\documentclass[11pt]{article}
\pdfoutput=1

\usepackage[utf8]{inputenc}
\usepackage{hyperref}
\usepackage[margin=0.8 in]{geometry}
\usepackage[UKenglish]{babel}
\usepackage[useregional]{datetime2}
\usepackage{graphicx}
\usepackage{amsmath, amssymb, mathtools, mathrsfs, dsfont}
\usepackage{enumerate}
\usepackage{authblk}
\graphicspath{/figures}

\renewcommand{\Re}{\operatorname{Re}}
\renewcommand{\Im}{\operatorname{Im}}

\newcommand{\hf}{\frac{1}{2}}
\newcommand{\req}[1]{(\ref{#1})}
\newcommand{\hyp}{{}_2F_1}

\newcommand{\dd}{\mathrm{d}}

\title{\textbf{Making near-extremal wormholes traversable}}
\author{Seamus Fallows\footnote{seamus.fallows@durham.ac.uk} }
\author{Simon F. Ross\footnote{s.f.ross@durham.ac.uk}}
\affil{\textit{Centre for Particle Theory, Department of Mathematical Sciences Durham University, South Road, Durham DH1 3LE, U.K.}}
\date{\today}

\begin{document}
\maketitle
\begin{abstract}
We construct a traversable wormhole from a charged AdS black hole by adding a coupling between the two boundary theories. We investigate how the effect of this deformation behaves in the extremal limit of the black hole. The black holes have finite entropy but an infinitely long throat in the extremal limit. We argue that it is still possible to make the throat traversable even in the extremal limit, but this requires either tuning the field for which we add a boundary coupling close to an instability threshold or scaling the strength of the coupling inversely with the temperature. In the latter case we show that the amount of information that can be sent through the wormhole scales with the entropy.  
\end{abstract}
\newpage
\tableofcontents 

\section{Introduction}

Time-independent black hole solutions have a wormhole, or Einstein-Rosen bridge, connecting two asymptotic regions. In holography, these solutions are related to entangled states in two copies of the dual CFT \cite{Maldacena:2001kr}. In the classical solution, this wormhole is not traversable; the two asymptotic regions are causally disconnected. In the holographic theory, this is a consequence of the fact that the two copies of the CFT are not coupled (only entangled), so no signal can propagate from one to the other. In \cite{Wall}, a simple coupling between the two boundaries was shown to make the wormhole traversable. In addition to realising the dreams of many science fiction authors, this provides a new insight into the relation between entanglement and spacetime in holographic theories: the passage of a bulk observer through the wormhole can be understood as quantum teleportation in the dual theory, using the entanglement of the dual state as a resource and using the coupling to communicate the needed classical information from one theory to the other \cite{Susskind:2017nto,Mald}. 

Much of the quantitative analysis of this phenomenon has focused on the simple example of the BTZ black hole in three dimensions, dual to a thermofield double (TFD) state in two copies of a two-dimensional CFT (although TFD states translated in time were considered in \cite{vanBreukelen:2017dul}, and rotating BTZ was considered in \cite{Caceres:2018ehr}). It is interesting to extend the discussion to more general cases: any entangled state can be used to realise quantum teleportation, but the bulk description in terms of a traversable wormhole may be special to particular forms of entanglement. 

In this paper, we take a step in this direction, by considering adding a boundary coupling to a charged Reissner-N\"ordstrom black hole in AdS$_{d+1}$, dual to a TFD state with a chemical potential for the charge in the CFT. The interest in this case is that the black holes have finite entropy (indicating finite entanglement in the dual state) but an infinitely long throat in the extremal limit. We would like to understand how difficult it is to make this infinite wormhole traversable, enabling communication between the two CFTs through the bulk. The divergence in the length of the throat implies that the correlation functions of operators on different boundaries vanishes in the extremal limit, unless the field dual to the operator is tuned to the threshold of an instability \cite{Andrade:2013rra}, suggesting that the effect of the boundary coupling on the bulk geometry may also vanish in this limit. Indeed, we find that unless we tune the bulk field to this instability threshold, we need to take the coupling between the two boundaries to scale to infinity as an inverse power of the temperature to have a finite effect on the bulk geometry in the extremal limit. If we accept this tuning of the coupling, however, we can communicate an amount of information that scales with the entropy of the black hole through the wormhole in the bulk. This is qualitatively different from the quotient construction of \cite{Fu:2018oaq,Fu:2019vco}, where the traversability of the wormhole traversable increased in the extremal limit. The key reason for this difference is that the double trace deformation we consider is a marginal or irrelevant deformation in the near-horizon AdS$_2$ region.  

In section \ref{geom}, we review the bulk solution, its extremal limit, and the dual CFT state. In section \ref{worm}, we add a double-trace boundary coupling and consider the resulting bulk deformation. There are no analytic solutions for the propagator of bulk fields on the full black hole background, so in our analysis we focus on the near-horizon region, which in the extremal limit has an $\mathrm{AdS}_2 \times S^{d-1}$ geometry. Boundary couplings on $\mathrm{AdS}_2$ and traversable wormholes have been considered previously \cite{Mald,Maldacena:2018lmt,Maldacena:2018gjk,Bak}, but our case is different as we emphasize the relation to the extremal limit of the asymptotic charged black hole geometry; we consider a charged field on the original near extremal black hole, which reduces to a charged field on AdS$_2$ with a uniform electric field background. We explicitly calculate the propagator for this charged field with the double-trace boundary condition. 

We find that to obtain a non-trivial opening of the wormhole, we need to either consider the operators dual to fields at threshold, or take the strength of the coupling to infinity as we take the temperature to zero. We argue that the limit of infinite coupling remains under control, precisely because the distance between the two boundaries in the bulk diverges, so the back-reaction in the bulk remains finite. Under these conditions, the coupling leads to a traversable wormhole in the bulk. The timescale for travel through this wormhole is set by the temperature of the black hole. 

We consider the back-reaction of a particle propagating through the wormhole in section \ref{bound}, and infer bounds on the amount of information that can be transmitted through the wormhole. We find that the bound is related to the entropy of the black hole, as expected from the relation to quantum teleportation. This indicates that this entropy from the entanglement of ground states is ``available'' as a resource for teleportation using simple boundary couplings, just as the thermal entropy in the usual TFD state was. 

It would be interesting to extend the calculations to consider other, general entangled states of the dual field theory where the two-point functions between the two boundaries are suppressed \cite{Marolf:2013dba,Balasubramanian:2014gla}. The entanglement in such states in principle provides a resource for quantum teleportation, but it is not clear if this teleportation could have a bulk description as in \cite{Wall}. It would also be interesting to consider states where the dual field theory has interacted with an environment, as in \cite{Verlinde:2020upt}. 

\section{Bulk geometry and boundary CFT}
\label{geom}

\subsection{RNAdS bulk solution}

We consider Einstein-Maxwell gravity with a negative cosmological constant. The action is
\begin{equation}
    S=\frac{1}{2\kappa^2}\int \dd^{d+1}x\sqrt{-g}\left[ (R-2\Lambda)-\frac{\ell^2}{g_F^2}F^2\right],
\end{equation}
where $g_F$ is an effective dimensionless gauge coupling and the cosmological constant is related to the AdS radius by
\begin{equation}
    \Lambda = -\frac{d(d-1)}{2\ell^2}.
\end{equation}
The theory admits a spherically symmetric Reissner-N\"ordstrom AdS black solution with the metric and gauge field given by 
\begin{equation}
    \dd s^2=-f(r)\dd t^2+\frac{\dd r^2}{f(r)}+r^2\dd\Omega_{d-1}^2, \qquad A =\mu \left(1-\frac{r_0^{d-2}}{r^{d-2}}\right)\dd t,
\end{equation}
where $\dd\Omega_{d-1}^2$ is the round metric on $S^{d-1}$ and
\begin{equation}
    f(r)\equiv 1-\frac{M}{r^{d-2}}+\frac{Q^2}{r^{2d-4}}+\frac{r^2}{\ell^2}.
\end{equation}
The full black hole geometry has two asymptotic regions, connected by an Einstein-Rosen bridge. These coordinates cover one of the asymptotic regions. 

The constants $Q$ and $M$ are proportional to the charge and ADM mass of the black hole respectively. The gauge field $A_t$ is dual to a conserved current $J_t$ in the boundary theory corresponding to a global $\mathrm{U}(1)$ symmetry. According to the holographic dictionary, the boundary value of the gauge field, $\mu=A_t(r\to\infty)$, is equal to the source of the conserved current, i.e. $\mu$ is the chemical potential in the field theory. It is related to the other bulk quantities through
\begin{equation}
    \mu=\sqrt{\frac{d-1}{2(d-2)}}\frac{g_FQ}{\ell r_0^{d-2}},
\end{equation}
where $r_0$ is the horizon radius, the largest positive root of the metric function $f(r_0)=0$. Far from the black hole horizon, $r\gg r_0$, the metric reduces to that of $\mathrm{AdS}_{d+1}$ in global coordinates. For fixed mass $M$ there is an open interval $Q\in (0,Q_*)$ for which $f$ has two distinct positive roots. As $Q$ approaches the extremal charge $Q_*$, these two roots converge and $f$ develops a double root at $r=r_*$. For $Q>Q_*$, the metric function has no positive roots and no black hole solution exists. The temperature of the black hole is 
\begin{equation}
    T = \frac{d-2}{4\pi r_0}\left[1+\frac{d r_0^2}{(d-2)\ell^2}-\frac{Q^2}{r_0^{2d-4}}\right]= \frac{d-2}{4\pi r_0}\left[1+\frac{d r_0^2}{(d-2)\ell^2}-\mu^2 \frac{2(d-2)\ell^2}{(d-1) g_F^2} \right].
\end{equation}
In the extremal limit $Q \to Q_*$, $T \to 0$. If we work in the grand canonical ensemble with fixed $\mu$, zero temperature is only reached if $\mu^2 > \mu_c^2 = \frac{(d-1)g_F^2}{2(d-2) \ell^2}$. 

The Euclidean black hole geometry is a saddle-point for the dual CFT in an appropriate ensemble, and the Lorentzian black hole is a saddle-point for the TFD state obtained by slicing the Euclidean path integral defining the ensemble in half. The TFD state for the grand canonical ensemble is \cite{Andrade:2013rra}
\begin{equation}
   |\psi \rangle = \frac{1}{\sqrt{Z}} \sum_i e^{-\beta (E_i +\mu Q_i)/2} |E_i, Q_i \rangle_1 \otimes |E_i, -Q_i \rangle_2.
\end{equation}
This is a state in the Hilbert space of two copies of the CFT, $|\psi \rangle \in \mathcal H_1 \otimes \mathcal H_2$, corresponding to the two asymptotic boundaries in the full spacetime, where $|E_i, Q_i \rangle$ are a basis of eigenstates of the Hamiltonian and the $U(1)$ charge in the CFT Hilbert space. For this state to be well-defined at low temperatures, $\beta \to \infty$, $E+\mu Q$ must be bounded below. The black hole is the dominant saddle-point in the grand canonical ensemble for all temperatures if $\mu > \mu_c$ \cite{Chamblin:1999tk}, so it provides the dual of this generalised TFD state. The finite entropy of the black hole in the extremal, zero-temperature limit implies an approximate degeneracy in the states at minimal $E+\mu Q$; in the extremal limit the TFD state remains entangled, with an entanglement entropy given by the black hole entropy.   

\subsection{Near horizon geometry}

In the zero temperature limit the metric develops a double pole at the horizon $r=r_*$. This implies that the black hole develops an infinite throat; the horizon is an infinite proper distance away on constant $t$ hypersurfaces. Taylor expanding, 
\begin{equation}
    f(r) =  \frac{1}{2} (r-r_{*})^2 f''(r_*) + \mathcal{O}(r-r_*)^3 \approx \frac{(r-r_*)^2}{\ell_2^2},
\end{equation}
where 
\begin{equation}
    \ell_2\equiv  \left[\frac{d(d-1)}{\ell^2}+\frac{(d-2)^2}{r_{*}^2} \right]^{-\hf}.
\end{equation}
For a large black hole $r_{*}\gg \ell$ we have $\ell_2\approx \ell/\sqrt{d(d-1)}$.  If we introduce the coordinate  
\begin{equation}
    \zeta=\frac{\ell_2^2}{r-r_{*}},
\end{equation}
the extremal geometry for large $\zeta$ is approximately $\mathrm{AdS}_2\times S^{d-1}$, 
\begin{equation}
    \dd s^2 \approx \frac{\ell_2^2}{\zeta^2}\left(-\dd t^2 + \dd\zeta^2\right)+r_{*}^2\dd\Omega_{d-1}^2, \qquad A \approx \frac{e_2}{\zeta}\dd t,
\end{equation}
where we have defined 
\begin{equation}
    e_2\equiv (d-2)\frac{\ell_2^2}{r_*}\mu_*=\sqrt{(d-1)(d-2)}\frac{\ell_2^2Q g_F}{\sqrt{2}\ell r_{*}^{d-1}},
\end{equation}
with $e_2\approx g_F/\sqrt{2d(d-1)}$ for $r_*\gg \ell$. We see that $\ell_2$ and $r_*$ become the radii of $\mathrm{AdS}_2$ and the $(d-1)$-sphere respectively. In this coordinate system the horizon is at $\zeta \to \infty$, and the geometry above is valid in a region of large $\zeta$, $\zeta > \zeta_c$ where $\zeta_c \sim \ell_2^2/r_*$ is a cutoff where we patch onto the full geometry, which is small for $r_* \gg \ell$. 

For near-extremal, finite temperature black holes, we in addition define 
\begin{equation}
    \zeta_0\equiv \frac{\ell_2^2}{r_0-r_{*}}.
\end{equation}
Close to extremality $\zeta_0 \gg \zeta_c$, and the near-horizon geometry becomes an $\mathrm{AdS}_2$ black hole,
\begin{equation}
    \dd s^2=\frac{\ell_2^2}{\zeta^2}\left[-\left(1-\frac{\zeta^2}{\zeta_0^2}\right) \dd t^2+\frac{\dd \zeta^2}{1-\frac{\zeta^2}{\zeta_0^2}}\right]+r_{*}^2\dd\Omega_{d-1}^2,  \qquad A=\frac{e_2}{\zeta}\left(1-\frac{\zeta}{\zeta_0}\right)\dd t,
\end{equation}
with inverse temperature $\beta=2\pi\zeta_0$. The extremal limit is $\zeta_0\to\infty$. Rescaling the coordinates $z=\zeta/\zeta_0$, $\tau=t/\zeta_0$, the AdS$_2$ metric becomes 
\begin{equation} \label{rindler}
    \dd s^2=\frac{\ell_2^2}{z^2}\left[-(1-z^2)\dd\tau^2+\frac{\dd z^2}{1-z^2}\right].
\end{equation}
with $\tilde{\beta}=2\pi$ and cut-off $z_c=\zeta_c/\zeta_0$. This is the metric of AdS$_2$ in Rindler coordinates. There is a horizon at $z=1$ and the conformal boundary is at $z=0$. We see that in these coordinates the extremal limit $\zeta_0\to\infty$ leaves the metric unchanged and acts to take the cut off $z_c\to 0$, reflecting the infinite length of the throat in the extremal limit. 

\begin{figure}[ht] 
  \centering
  \includegraphics[scale=0.7]{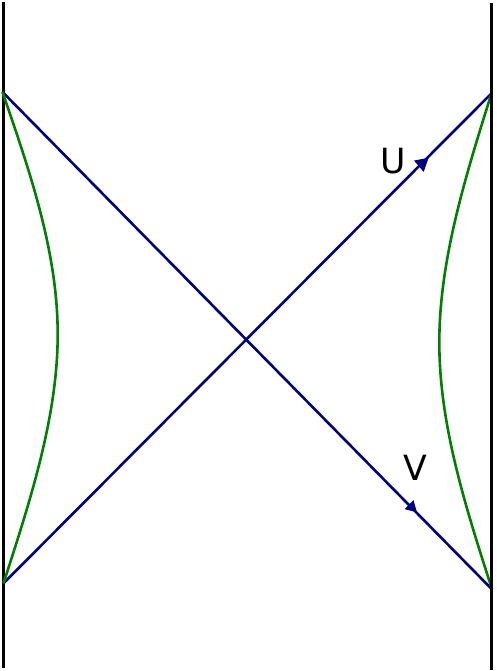}
  \caption{Near-horizon AdS$_2$ geometry of the near-extremal black hole, showing the cutoff boundaries and horizons. The coordinates $U,V$ increase towards the right boundary.}
  \label{Penrose}
\end{figure}

These Rindler coordinates cover the right wedge of the spacetime. To discuss the full AdS$_2$ black hole region, we will also work in Kruskal coordinates on the AdS$_2$, which are related to the Rindler coordinates above by 
\begin{equation} \label{kruskalc} 
U,V = \sqrt{\frac{1-z}{1+z}} e^{\pm \tau}. 
\end{equation}
In these coordinates, the metric and gauge field are 
\begin{equation} \label{kruskal}
    \dd s^2=\frac{4\ell_2^2\dd U\dd V}{(1-UV)^2}, \quad  A=e_2 \frac{V\dd U-U\dd V}{(1-UV)}.
\end{equation}
The bifurcation surface of the Rindler horizon is at $U=V=0$. The asymptotic boundaries are at $UV=1$; the right boundary has $U, V >0$ and the left boundary has $U, V <0$. The near-horizon geometry is pictured in figure \ref{Penrose}. 

It will also be useful later in discussing the back-reaction to write AdS$_2$ in terms of embedding coordinates $(X_0, X_1, X_2)$ in $\mathbb R^{2,1}$, where AdS$_2$ is realised as the universal cover of the hyperboloid $-X_0^2 - X_1^2 + X_2^2 = - \ell_2^2$. The embedding coordinates are related to Kruskal coordinates by 
\begin{equation}
    (X_0,X_1,X_2)= \ell_2 \left(\frac{U-V}{1-UV},\frac{1+UV}{1-UV},\frac{U+V}{1-UV}\right).
\end{equation}
If we define lightlike coordinates $X_{\pm} = X_0 \pm X_2$, the hyperboloid is $-X_+ X_- - X_1^2 = -\ell_2^2$, and 
\begin{equation} \label{embed} 
    (X_+,X_-,X_1)=\ell_2 \left(\frac{2U}{1-UV}, - \frac{2V}{1-UV}, \frac{1+UV}{1-UV} \right).
\end{equation}

The near-horizon $\mathrm{AdS}_2$ region is associated in the dual CFT description with a flow to a theory with an IR conformal symmetry acting just on the time direction \cite{Faulkner}. This IR conformal symmetry is broken by the deviation away from $\mathrm{AdS}_2$ in the full geometry, and we have a nearly AdS$_2$/ nearly CFT$_1$ duality in the IR \cite{Almheiri:2014cka,Maldacena:2016upp,Engelsoy:2016xyb}.\footnote{The one-dimensional conformal symmetry of the fixed point is not related to the conformal invariance of the ultraviolet $\mathrm{CFT}_d$ which is broken by the non-zero chemical potential.} The dynamics of the Einstein-Maxell theory in this near-horizon region of the RNAdS black hole reduces to JT gravity \cite{Nayak}.  

It is useful to organise bulk fields in the near-horizon region in terms of their scaling with respect to this IR conformal symmetry. Consider a bulk scalar field $\Phi(t,r,\Omega)$ of mass $m$ and charge $q$ on the full RNAdS background, dual to a local operator $\mathcal{O}(t,\Omega)$ in the UV boundary theory. 

Expanding in spherical harmonics on the sphere,
\begin{equation}
   \Phi(x)=\sum_{l,\mathbf{m}}\phi_{l\mathbf{m}}(t,\zeta)Y_{l\mathbf{m}}(\Omega), \qquad \int_{S^{d-1}_{r_*
}} \dd\Omega Y_{l\mathbf{m}}^*Y_{l'\mathbf{m}'}=\delta_{ll'}\delta_{\mathbf{m}\mathbf{m}'},
\end{equation}
the field modes $\phi_{l\mathbf{m}}$ are scalar fields on AdS$_2$ of mass 
\begin{equation}
    m_l^2\equiv m^2+\frac{l(l+d-2)}{r_{*}^2}.
\end{equation}
The coupling to the gauge field implies these fields are dual to operators of scaling dimension  \cite{Faulkner}
\begin{equation} \label{Delta}
    \Delta= \frac{1}{2}+\sqrt{\frac{1}{4}+m_l^2\ell_2^2 -q^2 e_2^2}.
\end{equation} 
If we take $q^2 e_2^2 > m^2\ell_2^2 + \frac{1}{4}$, the scalar field is unstable to condensing in the near-horizon AdS$_2$ region \cite{Gubser:2008px,Hartnoll:2008vx}, and the RNAdS solution will become unstable sufficiently close to extremality. We will be interested in studying fields just below this instability threshold, corresponding to $\Delta \simeq \frac{1}{2}$.\footnote{In AdS$_2$, we could obtain operators with $\Delta < \hf$ by considering the alternative quantization of the scalar field, but the near-horizon limit of the higher-dimensional solution gives us the standard quantization, so $\Delta \geq \hf$.}

\section{Wormhole construction}
\label{worm}

We want to consider the analogue of the traversable wormhole construction of \cite{Wall} for this black hole. This involves turning on  a  double  trace  deformation  coupling  the  two  CFTs  on  the  left  and  right  cut-off boundaries with a time-dependant Hamiltonian
\begin{equation} \label{source}
    \delta H(t,\zeta_c)=-h(t)\mathcal{O}_L(-t,\zeta_c)\mathcal{O}_R(t,\zeta_c),
\end{equation}
where $h(t)$ is a coupling which we take to vanish for $t<t_0$, and $\mathcal{O}$ is a boundary CFT operator dual to some bulk scalar field $\Phi$ on the RNAdS black hole. This coupling is dual to a modified boundary condition for the scalar $\Phi$ relating the fast fall-off part of the scalar at one asymptotic boundary to the slow fall-off part at the other and vice-versa. The idea of \cite{Wall} is that introducing this coupling (with an appropriate choice of sign of $h$) produces a quantum stress tensor which violates the averaged null energy condition (ANEC) along the black hole horizon. That is, $\int dU \langle T_{UU} \rangle < 0$, where $U$ is an affine parameter along the horizon. This ANEC violation means that the back-reaction of this quantum stress tensor can make the wormhole traversable; an observer crossing the horizon from one asymptotic region experiences a time advance due to the negative null energy (crossing the horizon moves them to an earlier time), and if they enter sufficiently early this enables them to escape into the other asymptotic region. 

In \cite{Wall}, this calculation was carried out on the BTZ black hole, where it was possible to calculate the propagator for the scalar field with the modified boundary condition explicitly, at leading order in the coupling $h$, and hence to obtain the ANEC violating stress tensor on the horizon. We cannot do such a calculation explicitly in the full RNAdS black hole geometry, as the scalar propagator on this geometry is not known in closed form. We therefore focus on the calculation in the near-horizon AdS$_2$ region. We can see the essential physics of the extremal limit in this near-horizon region. In particular, we can study how the calculation is affected by the diverging length of the Einstein-Rosen bridge. As discussed in the previous section, in the Rindler coordinates of  \eqref{rindler}, this divergence is reflected in the cutoff approaching the boundary of the AdS$_2$ space, $z_c \to 0$. 

We will consider one of the scalar modes $\phi_{l\mathbf{m}}$ on the AdS$_2$ space, and take a double-trace coupling of the form \eqref{source} on the cutoff boundary at $z=z_c$ in AdS$_2$. This is not precisely the same as taking this double-trace coupling on the boundary of the full AdS$_{d+1}$ spacetime, but we assume that in the limit of large black holes $r_* \gg \ell$, the renormalization group flow from the AdS boundary to the near-horizon region has a small effect. 

As in \cite{Wall}, we then want to calculate the modified propagator for a charged scalar field on AdS$_2$ with these boundary conditions. 
Using the evolution operator $U(t,t_0)=\mathcal{T}e^{-i\int_{t_0}^t\dd t\delta H(t,\zeta_c)}$ in the interaction picture the modified Wightman function is
\begin{equation}
    \langle\phi^{H}_R(t,\zeta)\phi^{H\dagger}_R(t',\zeta')\rangle=\langle U^{-1}(t,t_0)\phi^{I}_R(t,\zeta)U(t,t_0)U^{-1}(t',t_0)\phi^{I\dagger}_R(t',\zeta')U(t',t_0) \rangle.
\end{equation}
The superscripts $H$ and $I$ represent the Heisenberg and interaction picture respectively.  To leading order in $h$ this is (suppressing the $\zeta$ coordinate at intermediate steps and omitting $I$)
\begin{align} 
    G_{+}^h & \equiv -i\int_{t_{0}}^{t} \dd t_{1} h(t_{1})\langle[\mathcal{O}_{L}(-t_{1}) \mathcal{O}_{R}(t_{1}), \phi_{R}^\dagger(t)] \phi_{R}(t^{\prime})\rangle- i \int_{t_{0}}^{t^{\prime}} \dd t_{1} h(t_{1})\langle\phi_{R}^\dagger(t)[\mathcal{O}_{L}(-t_{1}) \mathcal{O}_{R}(t_{1}), \phi_{R}(t^{\prime})]\rangle \nonumber\\
    & \approx i \int_{t_{0}}^{t} \dd t_{1} h(t_{1})\langle\phi_{R}(t^{\prime}) \mathcal{O}_{L}(-t_{1})\rangle\langle[\phi_{R}^\dagger(t),\mathcal{O}_{R}(t_{1})]\rangle+i \int_{t_{0}}^{t'} \dd t_{1} h(t_{1})\langle\phi_{R}^\dagger(t) \mathcal{O}_{L}(-t_{1})\rangle\langle[\phi_{R}(t'),\mathcal{O}_{R}(t_{1}) ]\rangle \nonumber\\
    & = i \int_{t_{0}}^{t} \dd t_{1} h(t_{1})\langle\phi_{R}(t^{\prime}) \mathcal{O}_{R}^\dagger(-t_{1}+i\beta/2)\rangle\langle[\phi_{R}^\dagger(t),\mathcal{O}_{R}(t_{1})]\rangle \nonumber\\
    &=-\int_{t_{0}}^{t} \dd t_{1} h(t_{1})\mathcal{G}_+(t',\zeta';-t_1+i\beta/2,\zeta_c)\mathcal{G}_{ret}^\dagger(t,\zeta;t_1,\zeta_c)
\end{align}
where $\mathcal{G}_{+,ret}$ are the Wightman and retarded bulk-to-boundary propagators respectively, with the standard Dirichlet boundary conditions, and we have used analytic continuation to write $t_L = t_R + i \beta/2$. In the second line we used large $N$ factorization and causality $[\mathcal{O}_L,\phi_R]=0$. The second term in the second line is zero from $\langle\phi \phi\rangle=\langle\phi^\dagger\phi^\dagger\rangle=0$. 

This expression is written in terms of the $t,\zeta$ coordinates obtained from the near-horizon limit of the RNAdS black hole; to make the dependence on the extremal limit more explicit, it is useful to switch to the $\tau, z$ Rindler coordinates. We have  
\begin{equation} \label{rindprop}
    G_{+}^h=-\zeta_0^{1-2\Delta}\int_{\tau_0}^{\tau} \dd \tau_1 h(\tau_1)\mathcal{G}_{+}(\tau',z';-\tau_1+i\pi,z_c) \mathcal{G}_{ret}^\dagger(\tau,z;\tau_1,z_c),
\end{equation}
we see that this vanishes in the extremal limit $\zeta_0 \to \infty$, unless $\Delta = \frac{1}{2}$. The wormhole is becoming infinitely long in this limit, so the bulk-boundary two-point functions $\mathcal G$ go to zero as $\zeta_0^{-\Delta}$, and the effect of the change on the boundary conditions on the propagator between points in the interior of the geometry is going to zero. There is an exception for fields with $\Delta = \frac{1}{2}$, which correspond, as discussed at the end of the previous section, to scalars on the threshold of instability. For this case the effect remains finite in the extremal limit. 

This discussion is assuming fixed coupling $h(t_1)$. We can instead take it to scale with the inverse temperature $\beta$. This source function has dimension $1 - 2 \Delta$, so we can take the coupling to scale as
\begin{equation} \label{hscale}
    h(t_1)=h\left(\frac{2\pi}{\beta}\right)^{1-2\Delta}\theta\left(\frac{2\pi}{\beta}(t_1-t_0)\right)=h \zeta_0^{2\Delta-1}\theta(\tau_1-\tau_0),
\end{equation} 
where $h$ is a dimensionless constant.\footnote{This scaling of the coupling is introduced by hand to offset the behaviour of the propagator. There is an RG flow from the AdS$_{d+1}$ boundary to the AdS$_2$ boundary, but this is unaffected by the extremal limit, as the matching surface remains at finite distance from points in the outside region in the extremal limit.} The scaling of the prefactor will then cancel the $\zeta_0^{1-2\Delta}$ term in $G^h_+$, giving us a finite result for $\Delta > \frac{1}{2}$. This requires a diverging boundary coupling in the extremal limit, but we see explicitly from the bulk propagator calculation that this has only a finite effect in the bulk. 

We will be interested in evaluating $G_+^h$ for bulk points on the Killing horizon. In the Kruskal coordinates, this corresponds to $V= V'=0$ and some values $U, U'$. On the right boundary, the Kruskal coordinates $U_1, V_1$ are related to $\tau_1, z_c$ by \eqref{kruskalc}, which for small $z_c$ gives $1-U_1 V_1 \approx 2 z_c$. On the left boundary, we have $(U_L, V_L) = -(V_R, U_R)$. Thus the modified propagator is 
\begin{equation} \label{krusprop}
    G_{+}^h=-\int_{U_0}^{U} \frac{\dd U_1}{U_1} h \mathcal{G}_{+}(U',0;-V_1,-U_1) \mathcal{G}_{ret}^\dagger(U,0;U_1,V_1)
\end{equation}
with $1-U_1 V_1= 2 z_c$.\footnote{The scaling of the boundary coupling assumed in \eqref{hscale} cancels against the explicit dependence on $\zeta_0$ in the propagator, so even though the boundary coupling is growing in the extremal limit, the perturbative calculation of the propagator remains valid so long as the dimensionless constant $h$ is small.} 

\subsection{Charged scalar in AdS$_2$}

To calculate $G_+^h$ explicitly, we need to know the bulk-boundary propagators for a charged scalar field on AdS$_2$, with the standard Dirichlet boundary conditions. By symmetry, the propagator for a neutral scalar on AdS$_{d+1}$ is a function only of the invariant distance between the two points. On AdS$_2$, the bulk-bulk Green's function is (see e.g. \cite{Ammon:2015wua})
\begin{equation} \label{neutral}
 G(x,x') = C_{\Delta} \xi^\Delta {}_2 F_1 \left( \frac{\Delta}{2}, \frac{\Delta+1}{2}; \frac{2\Delta+1}{2}; \xi^2\right),  
\end{equation} 
\begin{equation}
    C_{\Delta}\equiv \frac{\Gamma(\Delta)}{2^\Delta\pi^{1/2}(2\Delta-1)\Gamma(\Delta-\hf)}, \qquad \Delta=\hf+\sqrt{\frac{1}{4}+m^2\ell_2^2},
\end{equation}
where we represent the bulk points in terms of their embedding coordinates $X, X'$, thinking of AdS$_2$ as the hyperboloid $-X_0^2-X_1^2+X_2^2 = -\ell_2^2$ in flat $\mathbb{R}^{2,1}$, and $\xi = - 1/X \cdot X'$ is an $\mathrm{SL}(2)$\footnote{We use the notation $\mathrm{SL(2)\equiv\mathrm{SL}(2,\mathbb{R})}$.} invariant related to the invariant distance between the two points. 

For a charged scalar, by contrast, the Green's function cannot be written purely as an $\mathrm{SL}(2)$ invariant function of the coordinates. This is because the gauge field is not invariant under $\mathrm{SL}(2)$ transformations, so the scalar equation of motion isn't either. However, as the field strength is invariant, the gauge field must only transform by some gauge transformation. The solution of the
scalar equation of motion will then be some phase times an $\mathrm{SL}(2)$ invariant function of the coordinates,  $G(x,x')=e^{iqe_2\Lambda(x,x')}P(\xi)$. We can determine $P(\xi)$ by solving for $G$ in the case where the source is at the bifurcation surface of the Rindler horizon, that is at $U'=V'=0$ in the Kruskal coordinates of \eqref{kruskal}. 
The scalar equation of motion is 
\begin{equation}
    D^\mu D_\mu \phi -m^2\phi=0,
\end{equation}
where $D_\mu=\partial_\mu-iqA_{\mu}$. In this case $\xi=z$, and we expect the solution to be independent of $\tau$ by time-translation symmetry. Taking $\phi=P(\xi)$, the equation of motion becomes 
\begin{equation} \label{P}
    \xi^2\partial_{\xi}\left(\left(1-\xi^2\right)\partial_{\xi}P\right)-\left(m^2\ell_2^2-q^2e_2^2\frac{1-\xi}{1+\xi}\right)P=0.
\end{equation}
Taking the solution that is normalizable at infinity we get
\begin{equation} 
    P(\xi)=C_{\Delta}\left(\frac{\xi}{1-\xi}\right)^{\Delta}\left(\frac{1+\xi}{1-\xi}\right)^{iqe_2}\hyp\left(\Delta+iqe_2,\Delta+iqe_2;2\Delta;\frac{2\xi}{\xi-1}\right),
\end{equation}
with $\Delta$ now given by expression \req{Delta}.
For $q=0$, this reduces to \eqref{neutral} by applying a transformation formula for the hypergeometric function. 

In the gauge we have chosen for $A$, the phase factor $\Lambda$ in the Green's function vanishes for a source on the bifurcation surface. To find $\Lambda$ for a general point on the horizon we can move the source along the horizon using an $\mathrm{SL}(2)$ transformation. A basis for the Lie algebra $\mathfrak{sl}(2)$ in terms of the embedding coordinates $(X_\pm,X_1)$ is
\begin{equation}
    Q^a=\hf\varepsilon^{abc}J_{bc},\qquad\qquad J_{ab}=X_a\frac{\partial}{\partial X^b}-X_b\frac{\partial}{\partial X^a}.
\end{equation}
Let us consider the killing vector 
\begin{equation}
    v=Q^{+}=X_{-}\partial_1-X_{1}\partial_{-}=\partial_U-V^2\partial_V,
\end{equation}
we see that on the horizon $v$ generates translations along the horizon, so it can be used to move the source off the bifurcation surface. Under the action of this vector field, the gauge field changes by $\mathcal{L}_{v}A=\dd V$.
This means that under an infinitesimal transformation $x'=x+\epsilon v$ the gauge transformation required to return $A$ to the form in \req{kruskal} is $\Lambda=-\epsilon V$. This suggests that for a source at position $U'$ on the horizon we should make the ansatz $G(x,x')=e^{iq\Lambda(U',V)}P(\xi)$. Plugging this into \req{P} one finds that this is a solution provided 
\begin{equation}
    \Lambda=\ln(U'V-1).
\end{equation}
Thus, the bulk-bulk propagator for a source on the horizon is 
\begin{equation} \label{bulk-prop}
    G(U',0,U,V)=(U'V-1)^{iq e_2}P(\xi),
\end{equation}
where for a source point on the horizon, 
\begin{equation} \label{bulk-prop}
    \xi = \frac{1-UV}{1+UV-2U'V}.
\end{equation}
For a single propagator, the phase factor is not physical; one can always choose a gauge to set it to zero. However, the calculation for $G_+^h$ involves a product of propagators from the left and right boundaries to the horizon and it's not possible to chose a gauge which sets both of the phase factors to zero. The relative phase between the propagators is physical.

We now want to obtain the bulk-boundary propagator between a point on the horizon and points on the left and right boundaries. For the bulk-boundary propagator $\mathcal G_+$ between the left boundary and the horizon, let us consider a point on the horizon at $U' >0$, so that the two points are spacelike separated. We can then simply take
\begin{equation} 
 \mathcal G_+(U', 0, -V_1, -U_1) =  b_m z_c^{-\Delta} G(U', 0, -V_1, -U_1), 
\end{equation}
where the constant $b_m$ relating the bulk-bulk and bulk-boundary propagators is
\begin{equation} \label{bm}
b_m=
\begin{cases}
2\Delta-1, & \Delta > \hf, \\
\frac{1}{2}, & \Delta = \hf.
\end{cases}
\end{equation}
In the limit as the bulk point approaches the boundary, $\xi \approx \frac{z_c}{1-U'V}$, so the propagator simplifies, as the hypergeometric function is simply one to leading order. Thus 
\begin{equation}
\mathcal{G}_{+}(U', 0, -V_1, -U_1)= b_mC_{\Delta}e^{-\pi qe_2}\left(\frac{1}{1+U'U_1}\right)^{\Delta-iq e_2}  .  
\end{equation}

\begin{figure}[ht] 
  \centering
  \includegraphics[scale=0.7]{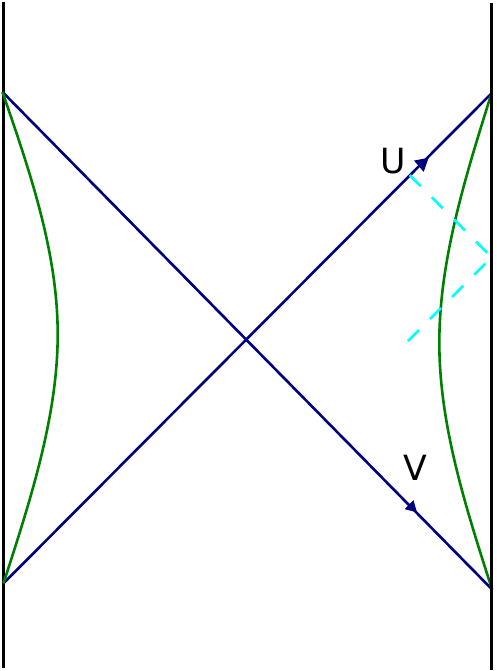}
  \caption{The portion of the cutoff boundary between the two dashed curves is connected to the point on the horizon by a geodesic.}
  \label{Geodesic}
\end{figure}

For the propagator to the right boundary, $\mathcal{G}_{ret}^\dagger(U,0;U_1,V_1)$, there is an interesting subtlety; at finite cutoff, there is a region of the boundary with $U_1 \in (U(1-2z_c),U)$ which is connected to the point $(U,0)$ by a timelike geodesic, as shown in figure \ref{Geodesic}. The structure of the propagator is different in this region. The size of this region goes to zero as $z_c \to 0$, but we need to check whether it makes a finite contribution to $G_+^h$. In this region, it is useful to make a change of variables $U_1=U (1-2z_c x)$, with $x \in (0,1)$. Then 
\begin{equation}
\frac{2\xi}{\xi-1} = \frac{(1-U_1V_1)}{V_1(U-U_1)} \approx   \frac{1}{x}.   
\end{equation}
Thus, the bulk-bulk propagator does not simplify in this region. However, $P(\xi)$ is a function only of $x$, with no dependence on $z_c$ at leading order, and the phase factor 
\begin{equation}
  (U V_1-1)^{iq e_2} \approx (2z_c (x-1))^{iq e_2}.
\end{equation}
The bulk-boundary propagator is thus $\mathcal{G}_{ret}^\dagger(U,0;U_1,V_1) = b_m z_c^{-\Delta} G(U, 0, U_1, V_1) \sim z_c^{-\Delta - iq e_2} f(x)$, so the contribution to $G_+^h$ from this region is 
\begin{equation} \label{krusprop}
    G_{+}^h \sim \int_{U (1-2 z_c)}^{U} \frac{\dd U_1}{U_1}  \mathcal{G}_{ret}^\dagger(U,0;U_1,V_1) \sim z_c^{1-\Delta - iq e_2}  \int_0^1 \dd x f(x), 
\end{equation}
so the contribution from this region vanishes in the limit as $z_c \to 0$ so long as $\Delta <1$. We will henceforth assume that we consider operators with $\frac{1}{2} \leq \Delta < 1$. 

In the region $U_1 \in (U_0, (1-2z_c) U)$, we have 
\begin{equation}
\frac{2\xi}{\xi-1} \approx \frac{z_c}{U V_1-1},   
\end{equation}
so 
\begin{equation}
    \mathcal{P} =b_m z_c^{-\Delta} P = b_m e^{-i\pi\Delta} (UV_1-1)^{-\Delta}, 
\end{equation}
and
\begin{equation}
    \mathcal{G}_{ret,R}=(UV_1-1)^{iqe_2}2\Im \mathcal P =-2b_m C_{\Delta} 
    \sin(\pi\Delta)\left(\frac{U_1}{U-U1}\right)^{\Delta-iqe_2}.
\end{equation}
We arrive at
\begin{equation} \label{G^h}
    G^h_{+} \approx -\frac{h\mathcal{C}_{\Delta}}{2}\int_{U_0}^{U}\frac{\dd U_1}{U_1}\left(\frac{1}{1+U_1 U'}\right)^{\Delta_q^*}\left(\frac{U_1}{U-U_1}\right)^{\Delta_q} \equiv -\frac{h\mathcal{C}_{\Delta}}{2} \int_{U_0}^{U}\frac{\dd U_1}{U_1} H(U,U',U_1),
\end{equation}
where $\Delta_q\equiv \Delta+iq e_2$ and $\mathcal{C}_{\Delta}\equiv 4b_m^2C_{\Delta}^2e^{-\pi qe_2}\sin(\pi\Delta)$. 
\subsection{Calculation of the stress tensor on the horizon}
We now calculate the quantum stress tensor on the horizon due to this boundary condition, showing that it leads to a violation of the ANEC. The stress tensor for a charged scalar is\footnote{As the matter fields are charged, they will source a change in the electric field as well, which changes the Maxwell stress tensor at the same order, but because of the index structure this does not contribute to the null-null component of the stress tensor we consider below.}
\begin{equation} \label{T}
    T_{\mu\nu}=(D_{\mu}\phi)(D_{\nu}\phi)^\dagger+(D_{\nu}\phi)(D_{\mu}\phi)^\dagger-g_{\mu\nu}g^{\rho\sigma}(D_{\rho}\phi)(D_{\sigma}\phi)^\dagger-g_{\mu\nu}m^2|\phi|^2.
\end{equation}
In the original AdS$_2$ geometry $g_{UU}=A_{U}=0$ on the horizon, so the terms involving the metric and gauge field drop out. The one-loop expectation value can then be related to the modified bulk propagator via point splitting,  
\begin{equation}
    \left\langle T_{UU}\right\rangle=2\langle\partial_{U}\phi\partial_{U}\phi^\dagger\rangle=\lim_{U'\to U}\partial_{U'}\partial_{U}(G^h_{+}(U,U')+G^{h\dagger}_{+}(U,U')).
\end{equation}
Evaluating the integral in \req{G^h} we find a closed form expression for the modified bulk propagator on the horizon\footnote{Note that we have used a transformation formula for the Appell function to get $G^h_+$ in this form.} 
\begin{equation}
G^h_{+}=\frac{h\mathcal{C}_\Delta}{2(\Delta_q-1)}\left(\frac{1}{1+U_0 U'}\right)^{\Delta_q^*}\left(\frac{U_0}{U-U_0}\right)^{\Delta_q-1} F_1\left(1;1-\Delta_q,\Delta_q^*;2-\Delta_q;1-\frac{U}{U_0},\frac{(U_0-U)U'}{1+U_0 U'}\right),
\end{equation}
where $F_1$ is the Appell hypergeometric function. Thus, from \req{T}, we also have a closed form expression for the quantum stress tensor.

The wormhole is rendered traversable if the ANEC is violated on the horizon, so we are interested in calculating the ANE given by
\begin{equation}
    \mathcal{A}^{\infty}(U_0)=\int_{U_0}^{\infty} \dd U \langle T_{UU}\rangle=2\int_{U_0}^{\infty}\dd U \lim_{U'\to U}\partial_{U'}\partial_U \Re G_{+}^h,
\end{equation}
where the superscript indicates that we are considering a source that is left on forever. Note that $\mathcal{A}^\infty$ has a simple relationship to the ANE for a source that is turned on for a finite interval $(U_0,U_f)$,
\begin{equation}
    \mathcal{A}(U_0,U_f)=\mathcal{A}^\infty(U_0)-\mathcal{A}^\infty(U_f).
\end{equation}
Rather than attempting the daunting task of directly integrating the stress tensor, we choose a different tack; instead, we consider an instantaneous source function given by
\begin{equation}
    h^{inst}(t_1)=h\left(\frac{2\pi}{\beta}\right)^{1-2\Delta}\delta\left(\frac{2\pi}{\beta}(t_1-t_0)\right)=\frac{h}{\zeta_0^{1-2\Delta}}U_0\delta(U_1-U_0).
\end{equation}
The ANE for this source is related to $\mathcal{A}^\infty$ by (see \cite{Frei})
\begin{equation}
    \mathcal{A}^{\infty}(U_0)=\int_{U_0}^{\infty}\frac{\dd u}{u}\mathcal{A}^{inst}(u),
\end{equation}
where the limits of integration are determined by $\mathcal{A}^\infty(\infty)=0$, i.e. if the source is never turned on, nothing happens. The delta function source significantly simplifies the calculation. For an instantaneous source, the modified bulk propagator is simply
\begin{equation}
    G^{inst}_+=-\hf h\mathcal{C}_{\Delta} H(U,U',U_0).
\end{equation}
To calculate the ANE, however, it is better to start with the general expression \eqref{G^h} and take the derivatives before setting the source to a delta function. This gives closed form expressions for both ANEs
\begin{align}
    \label{Ainst}
    \mathcal{A}^{inst}(U_0)&=\Re\left[\frac{h\mathcal{C}_{\Delta}\Gamma\left(1-\Delta_q\right)\Gamma\left(2\Re\Delta_q+1\right)}{\Gamma\left(\Delta_q^*\right)}\frac{U_0^{2\Delta_q+1}}{(1+U_0^2)^{2\Re\Delta_q+1}}\right], \\
    \mathcal{A}^{\infty}(U_0)&=\Re\left[\frac{h\mathcal{C}_{\Delta}\Gamma\left(1-\Delta_q\right)\Gamma\left(2\Delta_q+1\right)}{(2\Delta_q+1)\Gamma\left(\Delta_q\right)}\frac{\hyp\left(\hf+\Delta_q,\hf-\Delta_q;\frac{3}{2}+\Delta_q;\frac{1}{1+U_0^2}\right)}{(1+U_0^2)^{\Delta_q+\hf}} \right].
\end{align}

\begin{figure}[ht] 
  \centering
  \includegraphics[scale=0.7]{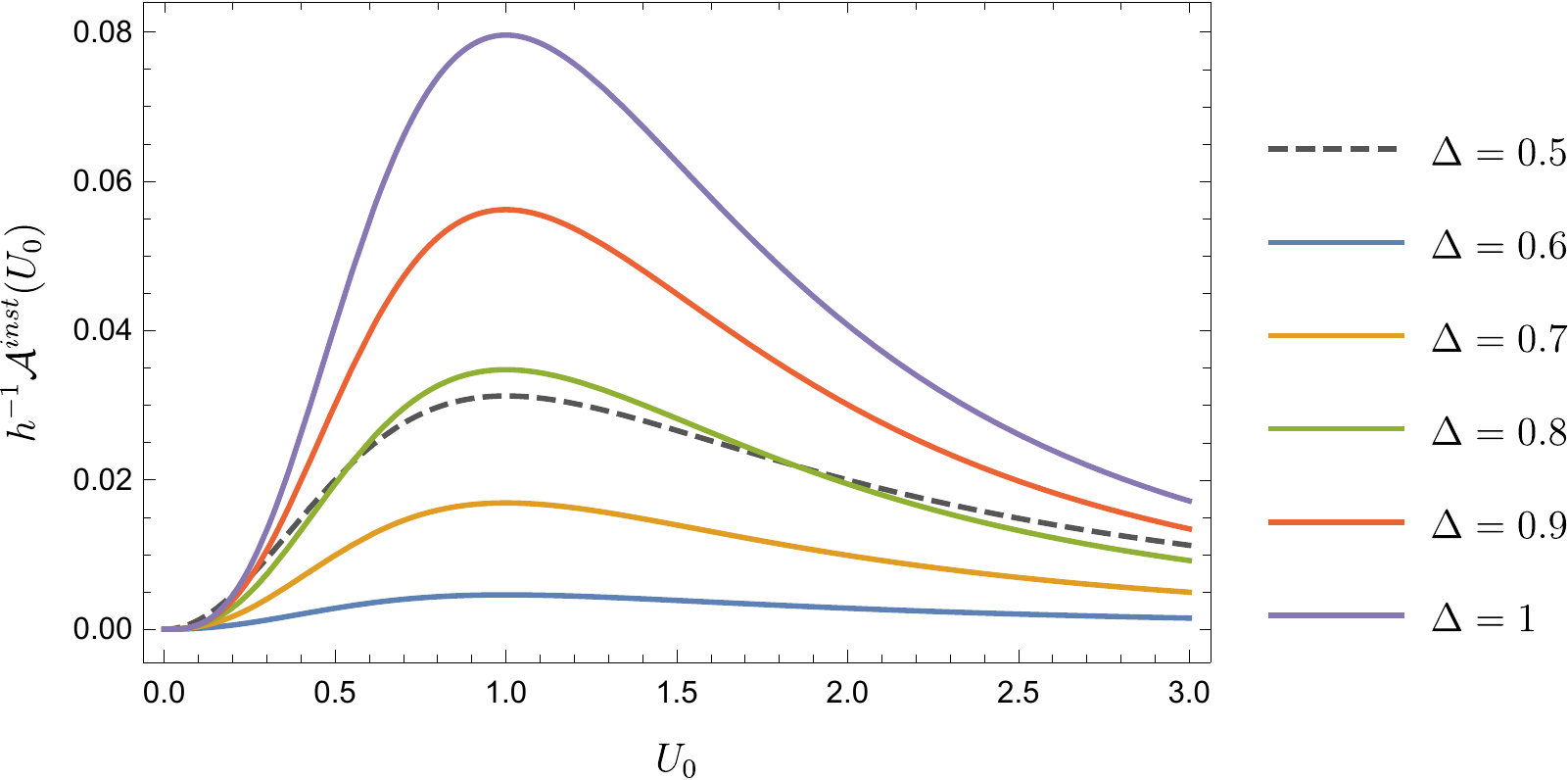}
  \includegraphics[scale=0.7]{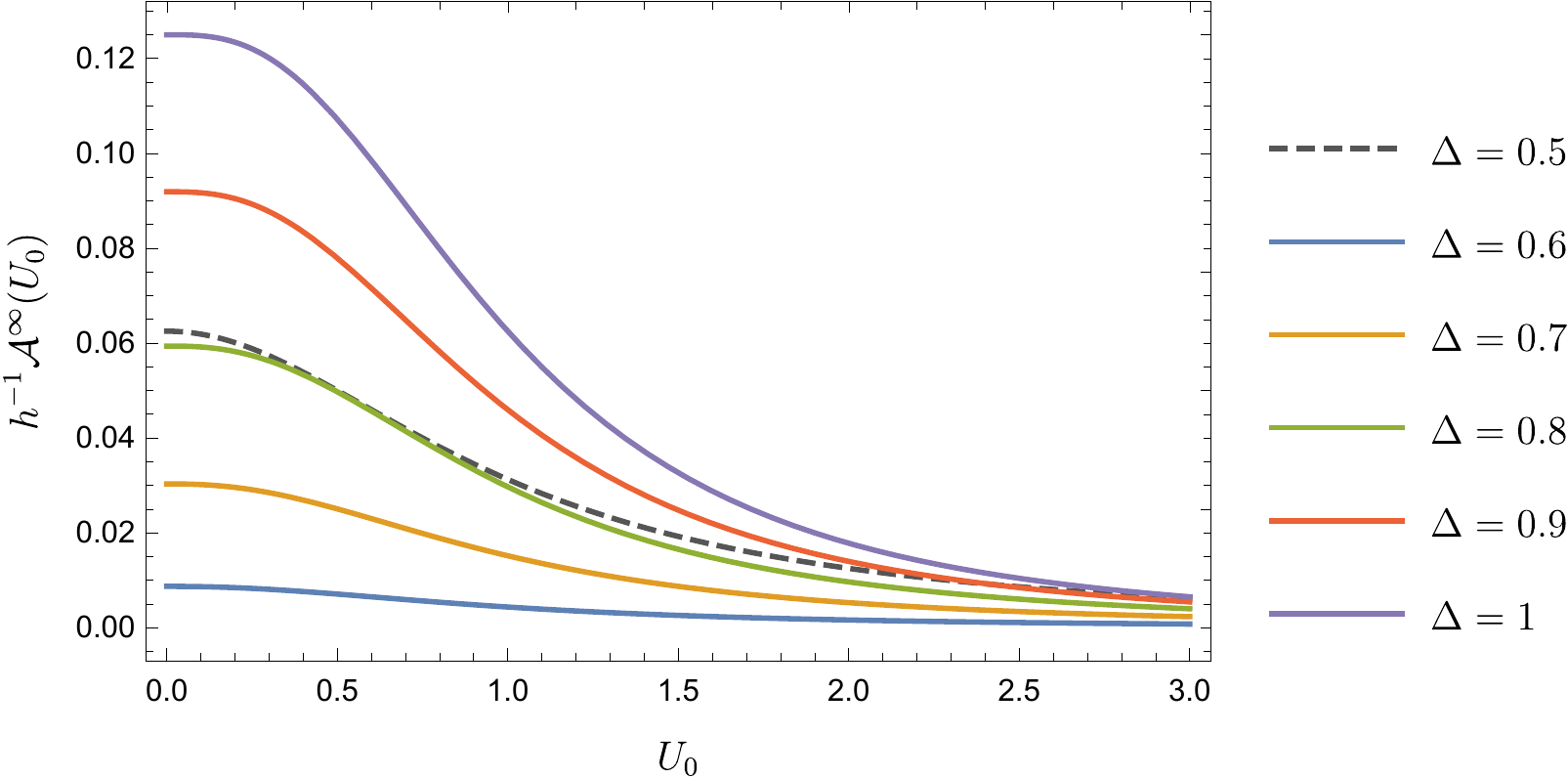}
  \caption{The ANE as a function of $U_0$ for $q=0$.}
  \label{ANE v U0, q=0}
\end{figure}

\begin{figure}[ht] 
  \centering
  \includegraphics[scale=0.7]{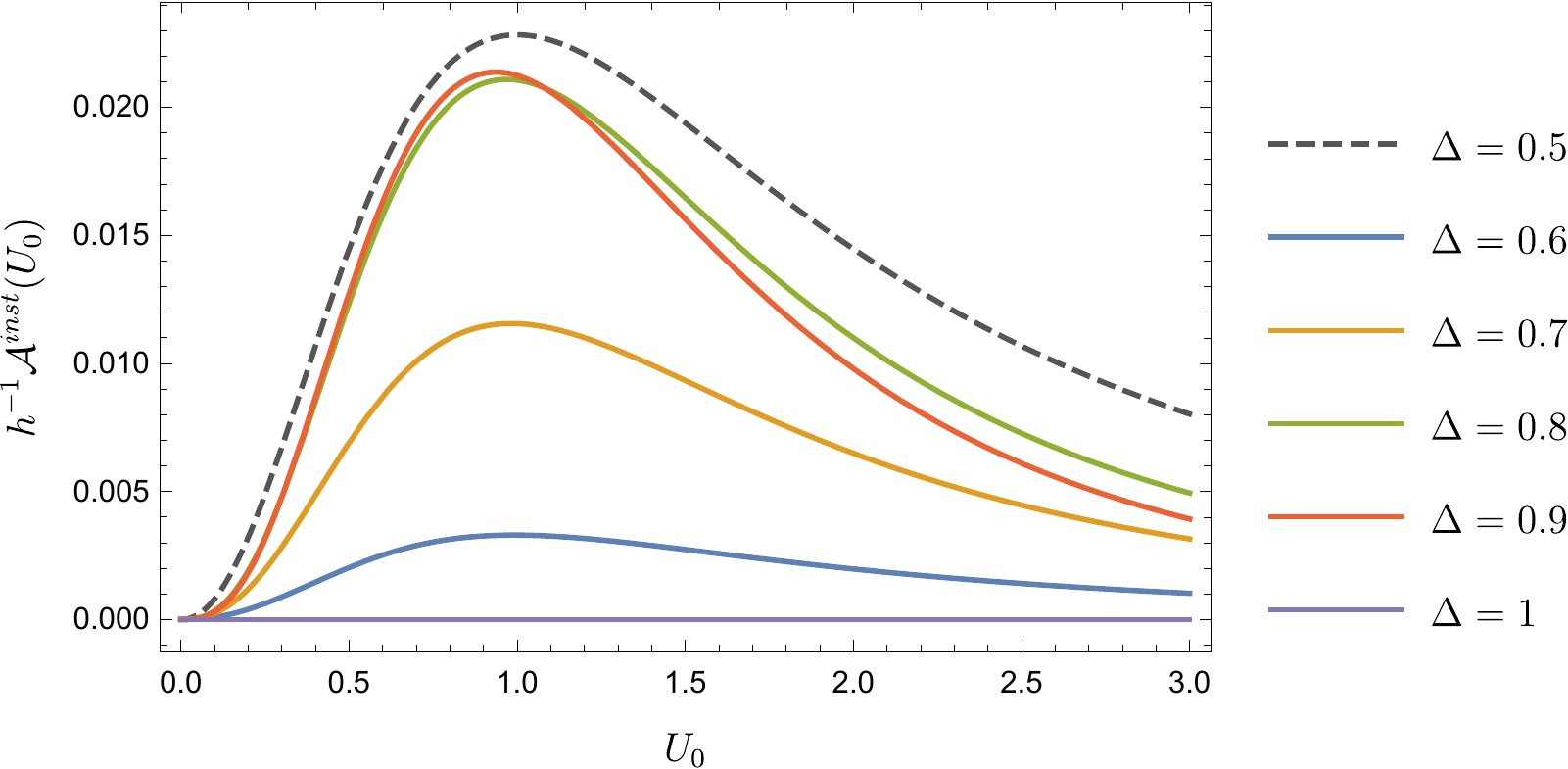}
  \includegraphics[scale=0.7]{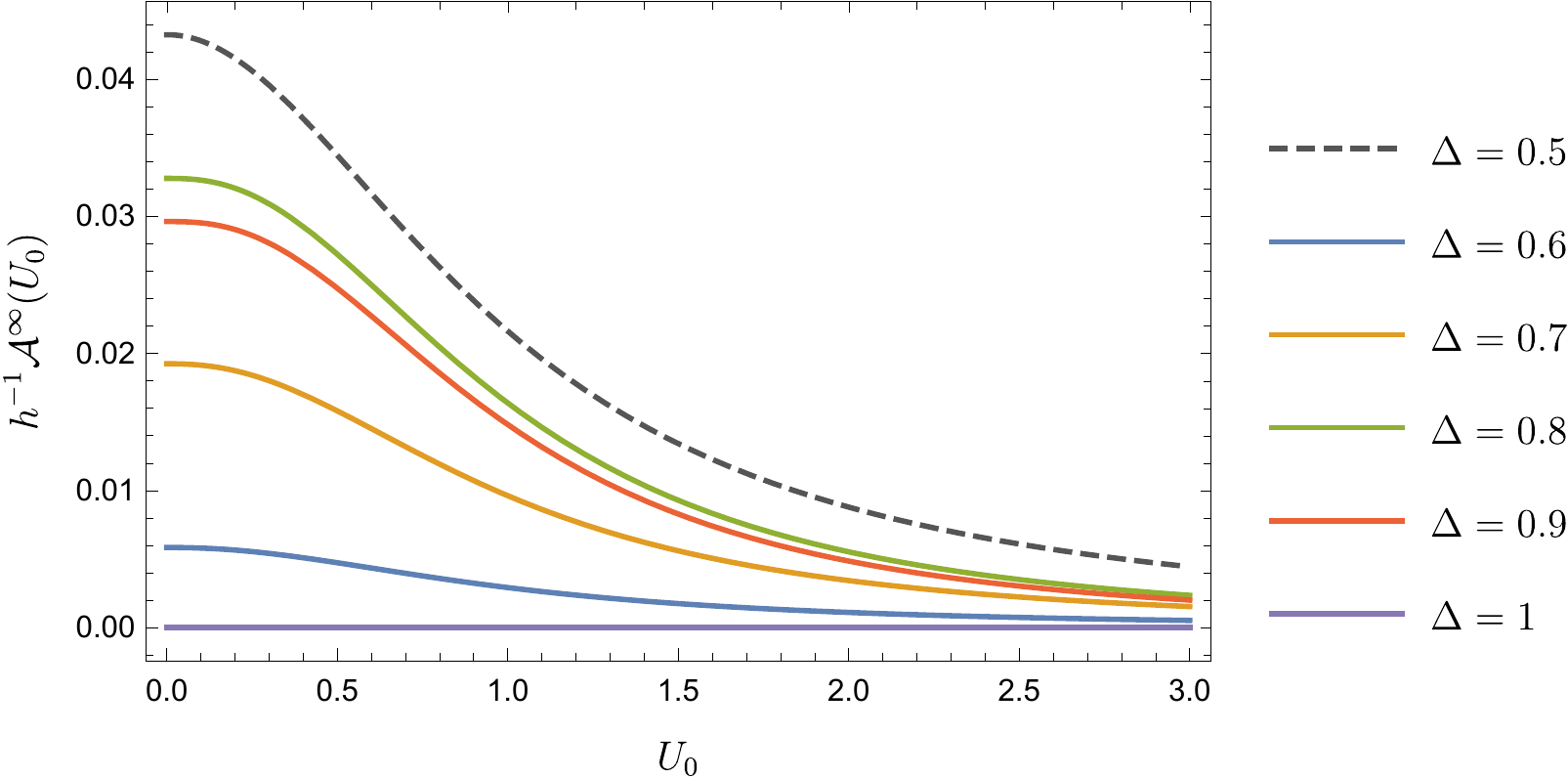}
  \caption{The ANE as a function of $U_0$ for non-zero $q e_2=0.1$.}
  \label{ANE v U0}
\end{figure}

It is interesting to note that if we considered an uncharged field, $q=0$, our final expression has the same $U_0$ dependence as was found for the BTZ black hole in \cite{Wall}. $\mathcal{A}^{inst}$ and $\mathcal{A}^\infty$ for $q=0$ are plotted against $U_0$ for different values of $\Delta$ in figure \ref{ANE v U0, q=0}. For the instantaneous source, maximal ANE is achieved when the non-local coupling is turned on at $U_0=1$ which corresponds to $t_R=t_L=0$. Conversely, $\mathcal{A}^\infty$ is maximal when the coupling is turned on in the infinite past $U_0=0$ ($t_R=t_L=-\infty$). For $q=0$ and $\Delta = \hf$, we have the simple expressions
\begin{align}
    \mathcal{A}^{inst}(U_0)&=\frac{h U_0^2}{8\left(1+U_0^2\right)^2} \\
    \mathcal{A}^\infty(U_0)&=\frac{h}{16\left(1+U_0^2\right)}.
\end{align}

\begin{figure}[ht] 
  \centering
  \includegraphics[scale=0.7]{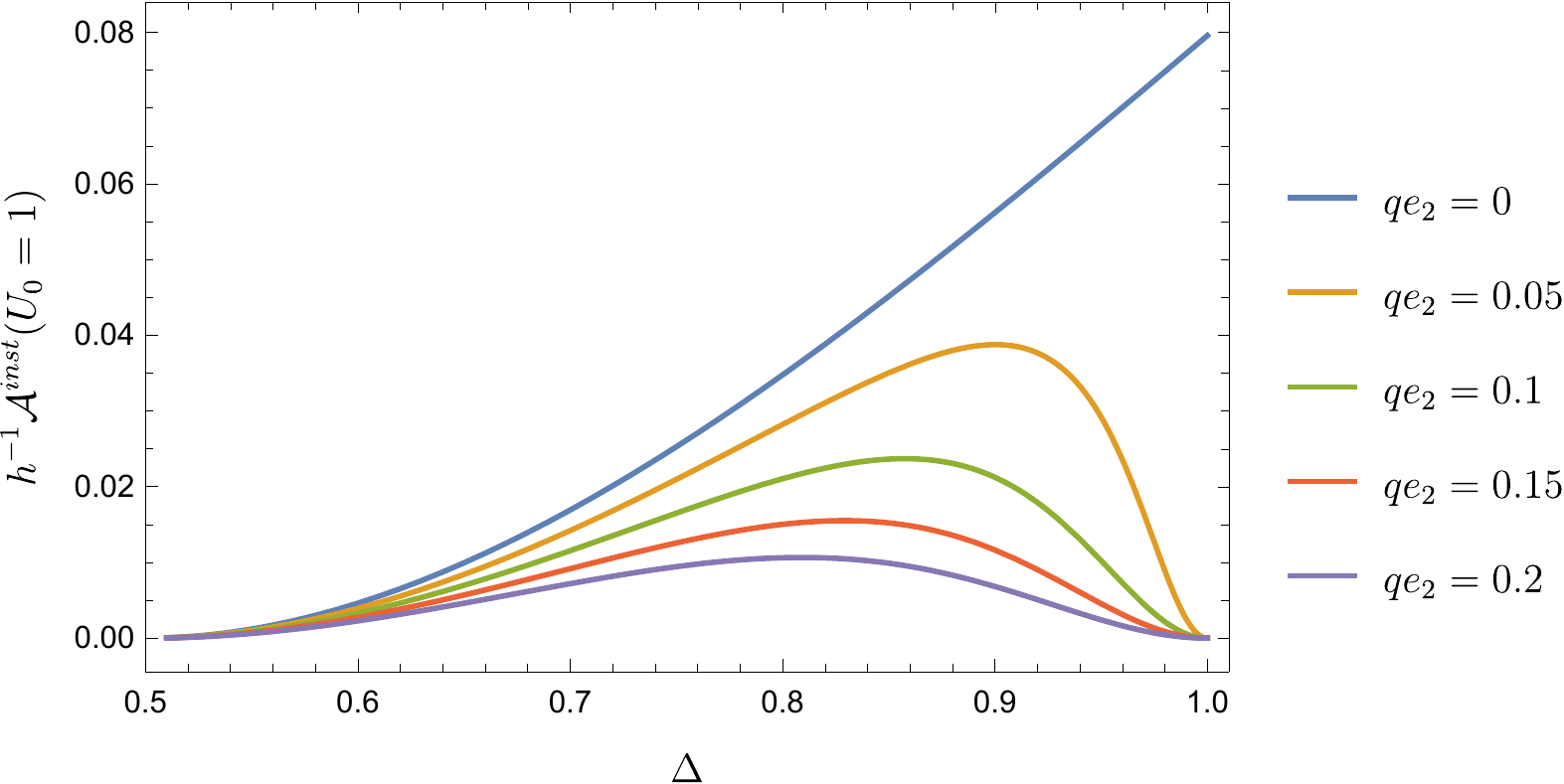}
  \includegraphics[scale=0.7]{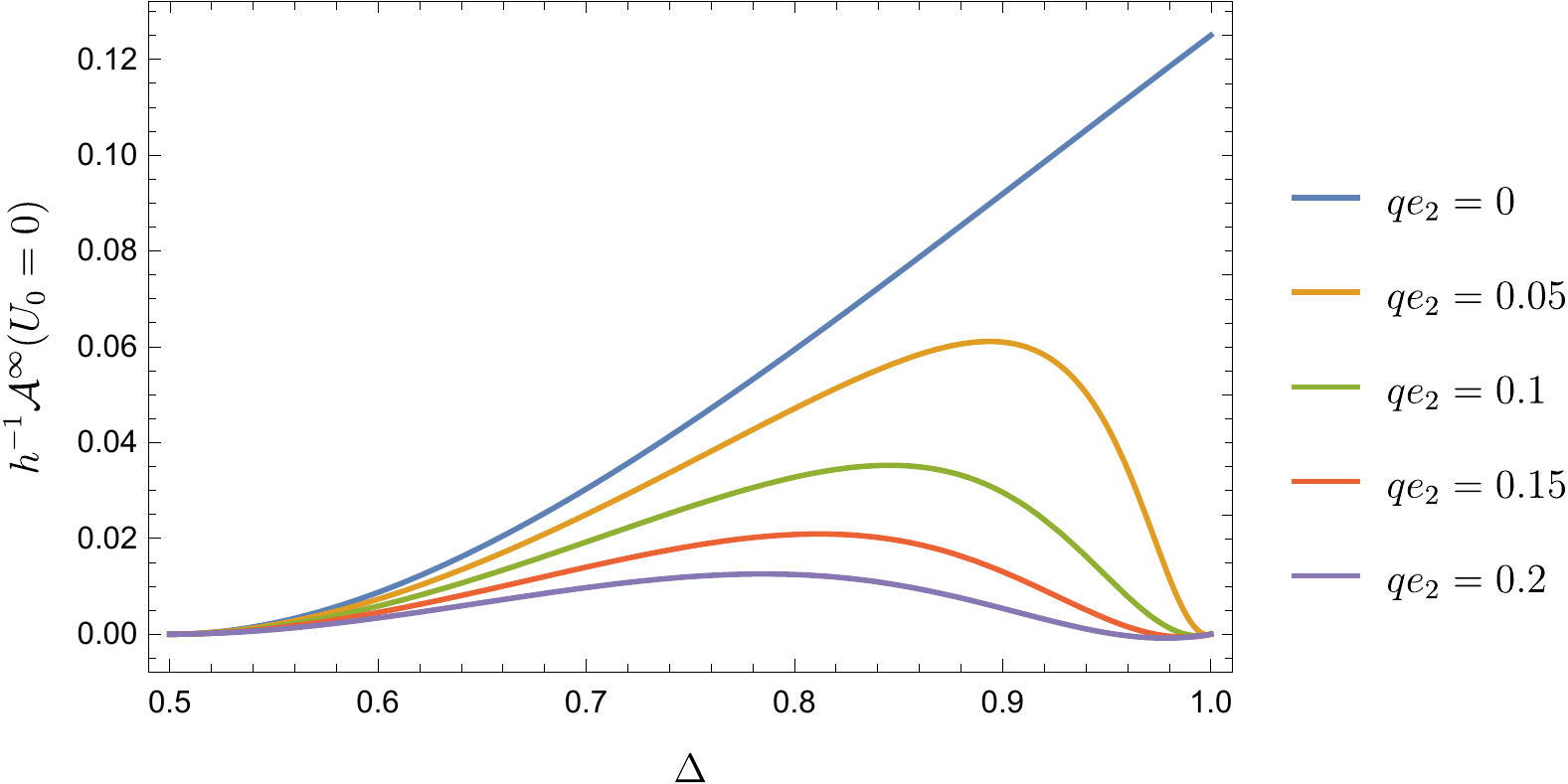}
  \caption{The ANE as a function of $\Delta$. For $q>0$, the ANE is suppressed at $\Delta=1$ by the factor of $\sin(\pi\Delta)$ in $\mathcal{C}_{\Delta}$. }
  \label{ANE v Delta}
\end{figure}

The ANE for $q>0$ is plotted against $U_0$ in figure \ref{ANE v U0}, and against $\Delta$ for representative values of $U_0$ in figure  \ref{ANE v Delta}. We see that while it increases with $\Delta$ for $q=0$, for $q>0$ there is some maximum at an intermediate value $\Delta\in(\hf,1)$. 

\section{Back-reaction and information bound}
\label{bound}

As discussed in the introduction, the back-reaction of this energy along the horizon will produce a time advance,  making it possible for a message from the left boundary sent in at early times to reach the right boundary. This makes the Einstein-Rosen bridge in the black hole into a traversable wormhole. We would like to understand how much information can be transmitted through the wormhole, which requires taking into account the back-reaction of the message. In the AdS$_2$ context, these back-reaction questions can be easily addressed using a JT gravity description of the nearly-AdS$_2$ gravitational dynamics, as in \cite{Mald} (see \cite{Maldacena:2016upp,Lin:2019qwu} for further discussion). The result is the same as in \cite{Mald}, as the back-reaction only depends on the ANE along the horizon, which have seen above is qualitatively the same for uncharged or charged fields. Thus, introducing a double-trace coupling for a single field only allows us to communicate order one bits of information from one boundary to the other. 

We will describe the calculation here briefly for completeness. In the JT gravity description, we take the bulk geometry to be fixed to be AdS$_2$, and the position of the boundaries is the dynamical information. When some matter is emitted into the bulk from one of the boundaries, the back-reaction causes the boundary trajectory to change. This change is described in terms of the $\mathrm{SL}(2)$ charges associated with the trajectories of the boundary and the emitted particles. 

In terms of the embedding coordinates $X^a$, the trajectories of the cutoff boundaries are described by $X \cdot Q = -2 \Phi_b$, where $Q^a$ is a vector in $\mathbb{R}^{2,1}$ which specifies the charges of the boundary trajectory under the $\mathrm{SL}(2)$ isometries of the bulk (we are thinking of the boundary as a particle moving in the bulk), and $\Phi_b$ is the boundary value of the dilaton. This equation gives a hyperbolic trajectory for the boundary, and the vector $Q^a$ can also be thought of as specifying the center of this hyperboloid $X^a=\bar{X}^a\propto Q^a$, that is, the point in the bulk which is light-like separated from the points where the trajectory meets the conformal boundary of AdS$_2$. For the near-horizon AdS$_2$ geometry described in \eqref{rindler}, the boundaries lie at $X_1 \approx \ell_2/z_c$, whose center is the bifurcation surface at $U=V=0$, that is $\bar X_\pm = 0, \bar{X}_1=\ell_2$. This implies $X \cdot \bar X = - \ell_2^2/z_c$. For the right boundary, the $\mathrm{SL}(2)$ charge is $Q_R = Q$, and for the left boundary, $Q_L = -Q$, so that the total $\mathrm{SL}(2)$ charge vanishes, $Q_L + Q_R =0$. 

If we inject matter into the bulk it will also carry an $\mathrm{SL}(2)$ charge. Matter particles in the bulk follow geodesics, which can be described by trajectories $X \cdot Q_m = 0$, where $Q_m$ is the $\mathrm{SL}(2)$ charge of the matter. The total $\mathrm{SL}(2)$ charge vanishes, $Q_L + Q_R + Q_m =0$, so the addition of matter will change the trajectories of the boundaries. If say the left boundary emits some positive energy matter, the recoil pushes it away, increasing the distance between the two boundaries. 

We are interested in two forms of back-reaction. First we consider the back-reaction of the bulk stress tensor due to the double-trace coupling. In the previous section, we calculated the null energy integrated along the horizon; this is precisely the charge 
\begin{equation} 
Q_{m,-} = \int \dd U \langle T_{UU} \rangle. 
\end{equation}
This matter was emitted by the right boundary, so this shifts the right boundary trajectory by $Q_R \to Q_R - Q_m$. The negative $Q_{m, -}$ thus moves the center of the right boundary trajectory to negative $V$; the shift $\Delta V = C Q_{m,-}/2$, where $C$ is a normalization factor depending on our conventions for the charges.  This makes it possible for messages leaving the left boundary at early times, at small negative $V$, to reach the right boundary. Note that the message needs to enter the wormhole at some finite time in the past in the Rindler time coordinate $\tau$; this implies that the time with respect to the asymptotic time $t = \zeta_0 \tau$ scales as the inverse temperature, so in the extremal limit the time it takes the message to go through the wormhole diverges. 

\begin{figure}[ht] 
  \centering
  \includegraphics[scale=1]{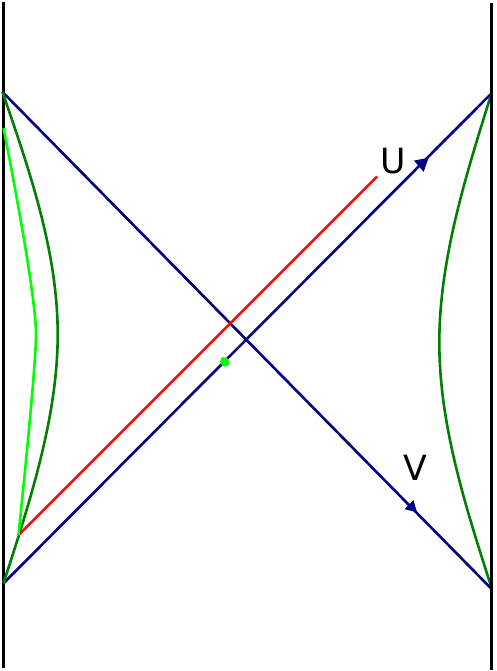}
  \caption{The emission of a null particle back-reacts on the trajectory of the left boundary, moving the center down along the horizon.}
  \label{Emission}
\end{figure}

Secondly, we consider the back-reaction on the left boundary of the emission of such a message. The message must be emitted at early times, so it is highly boosted relative to our bulk coordinate system, and will follow a nearly null trajectory in the bulk, with some momentum $Q'_{m, +} = p_V$. The back-reaction of the message shifts the left boundary by $Q_L \to Q_L - Q'_m$, moving the two boundaries further apart and suppressing the effect of the double-trace coupling. The center of the left boundary trajectory shifts down by $\delta U = C p_V/2$, as pictured in figure \ref{Emission}. This corresponds to transforming the boundary trajectory by a translation along the vector field $v$ considered earlier.  

This shift can thus be accounted for by a shift in the horizon coordinate in the calculation of the propagator from the left boundary to the horizon, so the integrand of the modified bulk propagator in the shockwave geometry is
\begin{equation}
    H_{\delta}(U,U',U_1)=\left(\frac{1}{1+U_1 (U'+\delta U)}\right)^{\Delta_q^*} \left(\frac{U_1}{U-U_1}\right)^{\Delta_q}.
\end{equation} 
Repeating the same analysis as before we find 
\begin{equation}
   \mathcal{A}^{inst}_{\delta}(U_0)=\Re\left[\frac{h\mathcal{C}_{\Delta}\Gamma\left(1-\Delta_q\right)\Gamma\left(2\Re\Delta_q+1\right)}{\Gamma\left(\Delta_q^*\right)}\frac{U_0^{2\Delta_q+1}}{\left(1+U_0^2\left(1+\frac{\delta U}{U_0}\right)\right)^{2\Re\Delta_q+1}}\right].
\end{equation}
Comparing this expression to \req{Ainst} we see that the probe approximation is valid for $\delta U/U_0\ll 1$. This implies that the total momentum carried by the message is bounded, 
\begin{equation} 
 \frac{p_V^{total}}{2} < \frac{U_0}{C}.
\end{equation}

A lower bound on the momentum carried by the individual particles making up the message can be found using the uncertainty principle, 
\begin{equation}
    p^{each}_V\gtrsim\frac{1}{\Delta V} = \frac{2 }{C |\mathcal{A}|}.
\end{equation}
Combining this with the probe approximation gives a bound on the number of bits that can be sent through the wormhole,
\begin{equation}
    N = \frac{p^{total}_V}{p^{each}_V} <   U_0 | \mathcal{A}|.
\end{equation}
We see from the discussion of the ANE in the previous section that the RHS takes values less than one. 

Thus, coupling a single field in the AdS$_2$ region would only allow us to send less than one bit of information before the back-reaction of the message starts to close up the wormhole. It might seem surprising that this result is independent of the entropy of the black hole; but this is just because we have focused on coupling a particular spherical harmonic $\phi_{\ell \mathbf{m}}$ of a $(d+1)$-dimensional scalar field $\Phi$. If we want to restrict attention to operators with $\Delta = \hf$, for which we can generate a traversable wormhole with a finite boundary coupling even in the extremal limit, we will only be able to consider the $s$-wave excitation of a scalar that saturates the instability threshold, and we will only be able to communicate less than a single bit for each field. However, if we allow consideration of operators with $\hf < \Delta < 1$ in the AdS$_2$ region, with a coupling that scales with the temperature, we get to consider a large number of spherical harmonics on the $S^{d-1}$: for $r_* \gg \ell$, we have $K$ spherical harmonics with $\Delta <1$ where
\begin{equation}
   K \sim \frac{r_*^{d-1}}{\ell_2^{d-1}} \sim \frac{A}{\ell^{d-1}}, 
\end{equation}
so the number of fields we can introduce such a coupling for, and hence the number of bits we can send through the wormhole, scales as the area of the horizon in AdS units, as in the BTZ analysis of \cite{Frei}. As in \cite{Frei}, to make the number of bits scale like the area in Planck units, we would need to consider a large number of $(d+1)$-dimensional fields $\Phi$.

\section*{Acknowledgements}

We thank Nabil Iqbal for useful discussions.
SFR is supported in part by STFC through grant ST/P000371/1, and SF is supported by an STFC studentship.

\bibliographystyle{JHEP}
\bibliography{wormhole}

\providecommand{\href}[2]{#2}\begingroup\raggedright\begin{thebibliography}{10}

\bibitem{Maldacena:2001kr}
J.~M. Maldacena, \emph{{Eternal black holes in anti-de Sitter}},
  \href{http://dx.doi.org/10.1088/1126-6708/2003/04/021}{\emph{JHEP} {\bf 04}
  (2003) 021}, [\href{https://arxiv.org/abs/hep-th/0106112}{{\tt
  hep-th/0106112}}].

\bibitem{Wall}
P.~Gao, D.~L. Jafferis and A.~C. Wall, \emph{{Traversable Wormholes via a
  Double Trace Deformation}},
  \href{http://dx.doi.org/10.1007/JHEP12(2017)151}{\emph{JHEP} {\bf 12} (2017)
  151}, [\href{https://arxiv.org/abs/1608.05687}{{\tt 1608.05687}}].

\bibitem{Susskind:2017nto}
L.~Susskind and Y.~Zhao, \emph{{Teleportation through the wormhole}},
  \href{http://dx.doi.org/10.1103/PhysRevD.98.046016}{\emph{Phys. Rev. D} {\bf
  98} (2018) 046016}, [\href{https://arxiv.org/abs/1707.04354}{{\tt
  1707.04354}}].

\bibitem{Mald}
J.~Maldacena, D.~Stanford and Z.~Yang, \emph{{Diving into traversable
  wormholes}}, \href{http://dx.doi.org/10.1002/prop.201700034}{\emph{Fortsch.
  Phys.} {\bf 65} (2017) 1700034},
  [\href{https://arxiv.org/abs/1704.05333}{{\tt 1704.05333}}].

\bibitem{vanBreukelen:2017dul}
R.~van Breukelen and K.~Papadodimas, \emph{{Quantum teleportation through
  time-shifted AdS wormholes}},
  \href{http://dx.doi.org/10.1007/JHEP08(2018)142}{\emph{JHEP} {\bf 08} (2018)
  142}, [\href{https://arxiv.org/abs/1708.09370}{{\tt 1708.09370}}].

\bibitem{Caceres:2018ehr}
E.~Caceres, A.~S. Misobuchi and M.-L. Xiao, \emph{{Rotating traversable
  wormholes in AdS}},
  \href{http://dx.doi.org/10.1007/JHEP12(2018)005}{\emph{JHEP} {\bf 12} (2018)
  005}, [\href{https://arxiv.org/abs/1807.07239}{{\tt 1807.07239}}].

\bibitem{Andrade:2013rra}
T.~Andrade, S.~Fischetti, D.~Marolf, S.~F. Ross and M.~Rozali,
  \emph{{Entanglement and correlations near extremality: CFTs dual to
  Reissner-Nordström $AdS_5$}},
  \href{http://dx.doi.org/10.1007/JHEP04(2014)023}{\emph{JHEP} {\bf 04} (2014)
  023}, [\href{https://arxiv.org/abs/1312.2839}{{\tt 1312.2839}}].

\bibitem{Fu:2018oaq}
Z.~Fu, B.~Grado-White and D.~Marolf, \emph{{A perturbative perspective on
  self-supporting wormholes}},
  \href{http://dx.doi.org/10.1088/1361-6382/aafcea}{\emph{Class. Quant. Grav.}
  {\bf 36} (2019) 045006}, [\href{https://arxiv.org/abs/1807.07917}{{\tt
  1807.07917}}].

\bibitem{Fu:2019vco}
Z.~Fu, B.~Grado-White and D.~Marolf, \emph{{Traversable Asymptotically Flat
  Wormholes with Short Transit Times}},
  \href{http://dx.doi.org/10.1088/1361-6382/ab56e4}{\emph{Class. Quant. Grav.}
  {\bf 36} (2019) 245018}, [\href{https://arxiv.org/abs/1908.03273}{{\tt
  1908.03273}}].

\bibitem{Maldacena:2018lmt}
J.~Maldacena and X.-L. Qi, \emph{{Eternal traversable wormhole}},
  \href{https://arxiv.org/abs/1804.00491}{{\tt 1804.00491}}.

\bibitem{Maldacena:2018gjk}
J.~Maldacena, A.~Milekhin and F.~Popov, \emph{{Traversable wormholes in four
  dimensions}},  \href{https://arxiv.org/abs/1807.04726}{{\tt 1807.04726}}.

\bibitem{Bak}
D.~Bak, C.~Kim and S.-H. Yi, \emph{{Bulk view of teleportation and traversable
  wormholes}}, \href{http://dx.doi.org/10.1007/JHEP08(2018)140}{\emph{JHEP}
  {\bf 08} (2018) 140}, [\href{https://arxiv.org/abs/1805.12349}{{\tt
  1805.12349}}].

\bibitem{Marolf:2013dba}
D.~Marolf and J.~Polchinski, \emph{{Gauge/Gravity Duality and the Black Hole
  Interior}},
  \href{http://dx.doi.org/10.1103/PhysRevLett.111.171301}{\emph{Phys. Rev.
  Lett.} {\bf 111} (2013) 171301}, [\href{https://arxiv.org/abs/1307.4706}{{\tt
  1307.4706}}].

\bibitem{Balasubramanian:2014gla}
V.~Balasubramanian, M.~Berkooz, S.~F. Ross and J.~Simon, \emph{{Black Holes,
  Entanglement and Random Matrices}},
  \href{http://dx.doi.org/10.1088/0264-9381/31/18/185009}{\emph{Class. Quant.
  Grav.} {\bf 31} (2014) 185009}, [\href{https://arxiv.org/abs/1404.6198}{{\tt
  1404.6198}}].

\bibitem{Verlinde:2020upt}
H.~Verlinde, \emph{{ER = EPR revisited: On the Entropy of an Einstein-Rosen
  Bridge}},  \href{https://arxiv.org/abs/2003.13117}{{\tt 2003.13117}}.

\bibitem{Chamblin:1999tk}
A.~Chamblin, R.~Emparan, C.~V. Johnson and R.~C. Myers, \emph{{Charged AdS
  black holes and catastrophic holography}},
  \href{http://dx.doi.org/10.1103/PhysRevD.60.064018}{\emph{Phys. Rev. D} {\bf
  60} (1999) 064018}, [\href{https://arxiv.org/abs/hep-th/9902170}{{\tt
  hep-th/9902170}}].

\bibitem{Faulkner}
T.~Faulkner, H.~Liu, J.~McGreevy and D.~Vegh, \emph{{Emergent quantum
  criticality, Fermi surfaces, and AdS(2)}},
  \href{http://dx.doi.org/10.1103/PhysRevD.83.125002}{\emph{Phys. Rev. D} {\bf
  83} (2011) 125002}, [\href{https://arxiv.org/abs/0907.2694}{{\tt
  0907.2694}}].

\bibitem{Almheiri:2014cka}
A.~Almheiri and J.~Polchinski, \emph{{Models of AdS$_{2}$ backreaction and
  holography}}, \href{http://dx.doi.org/10.1007/JHEP11(2015)014}{\emph{JHEP}
  {\bf 11} (2015) 014}, [\href{https://arxiv.org/abs/1402.6334}{{\tt
  1402.6334}}].

\bibitem{Maldacena:2016upp}
J.~Maldacena, D.~Stanford and Z.~Yang, \emph{{Conformal symmetry and its
  breaking in two dimensional Nearly Anti-de-Sitter space}},
  \href{http://dx.doi.org/10.1093/ptep/ptw124}{\emph{PTEP} {\bf 2016} (2016)
  12C104}, [\href{https://arxiv.org/abs/1606.01857}{{\tt 1606.01857}}].

\bibitem{Engelsoy:2016xyb}
J.~Engelsöy, T.~G. Mertens and H.~Verlinde, \emph{{An investigation of
  AdS$_{2}$ backreaction and holography}},
  \href{http://dx.doi.org/10.1007/JHEP07(2016)139}{\emph{JHEP} {\bf 07} (2016)
  139}, [\href{https://arxiv.org/abs/1606.03438}{{\tt 1606.03438}}].

\bibitem{Nayak}
P.~Nayak, A.~Shukla, R.~M. Soni, S.~P. Trivedi and V.~Vishal, \emph{{On the
  Dynamics of Near-Extremal Black Holes}},
  \href{http://dx.doi.org/10.1007/JHEP09(2018)048}{\emph{JHEP} {\bf 09} (2018)
  048}, [\href{https://arxiv.org/abs/1802.09547}{{\tt 1802.09547}}].

\bibitem{Gubser:2008px}
S.~S. Gubser, \emph{{Breaking an Abelian gauge symmetry near a black hole
  horizon}}, \href{http://dx.doi.org/10.1103/PhysRevD.78.065034}{\emph{Phys.
  Rev. D} {\bf 78} (2008) 065034}, [\href{https://arxiv.org/abs/0801.2977}{{\tt
  0801.2977}}].

\bibitem{Hartnoll:2008vx}
S.~A. Hartnoll, C.~P. Herzog and G.~T. Horowitz, \emph{{Building a Holographic
  Superconductor}},
  \href{http://dx.doi.org/10.1103/PhysRevLett.101.031601}{\emph{Phys. Rev.
  Lett.} {\bf 101} (2008) 031601}, [\href{https://arxiv.org/abs/0803.3295}{{\tt
  0803.3295}}].

\bibitem{Ammon:2015wua}
M.~Ammon and J.~Erdmenger, \emph{{Gauge/gravity duality}: {Foundations and
  applications}}.
\newblock Cambridge University Press, Cambridge, 4, 2015.

\bibitem{Frei}
B.~Freivogel, D.~A. Galante, D.~Nikolakopoulou and A.~Rotundo,
  \emph{{Traversable wormholes in AdS and bounds on information transfer}},
  \href{http://dx.doi.org/10.1007/JHEP01(2020)050}{\emph{JHEP} {\bf 01} (2020)
  050}, [\href{https://arxiv.org/abs/1907.13140}{{\tt 1907.13140}}].

\bibitem{Lin:2019qwu}
H.~W. Lin, J.~Maldacena and Y.~Zhao, \emph{{Symmetries Near the Horizon}},
  \href{http://dx.doi.org/10.1007/JHEP08(2019)049}{\emph{JHEP} {\bf 08} (2019)
  049}, [\href{https://arxiv.org/abs/1904.12820}{{\tt 1904.12820}}].

\end{thebibliography}\endgroup


\providecommand{\href}[2]{#2}\begingroup\raggedright\endgroup

\end{document}